%% file: main.tex
\pdfoutput=1
\documentclass[10pt, conference, compsocconf]{IEEEtran}
\usepackage{booktabs}
\usepackage{pbox}
\usepackage{amssymb}
\usepackage{url}
\usepackage{graphicx}
\usepackage{epstopdf}
\makeatletter
\let\MYcaption\@makecaption
\makeatother
\usepackage[font=footnotesize]{subcaption}
\makeatletter
\let\@makecaption\MYcaption
\makeatother
\usepackage{pgffor}
\usepackage{amsmath}
\usepackage{adjustbox}
\let\labelindent\relax
\usepackage{enumitem}
\setlist{leftmargin=5mm}
\usepackage{array}
\usepackage{setspace}
\usepackage{makecell}
\usepackage{pgfplots}
\usepackage{soul}
\usepackage[breakable, theorems, skins]{tcolorbox}
\usepackage{marvosym}
\usepackage[flushleft]{threeparttable}
\pgfplotsset{compat=1.3}
\usepackage{color, colortbl}
\usepackage{tikz}
\usepackage{pgf}
\usepackage{balance}
\usepackage{arydshln}
\usetikzlibrary{positioning}
\usepackage[linesnumbered,ruled,vlined]{algorithm2e}
\usetikzlibrary{arrows}
\usepackage[flushleft]{threeparttable}
\graphicspath{{images/} }

\sethlcolor{lightgray}

\newcolumntype{R}[2]{%
    >{\adjustbox{angle=#1,lap=\width-(#2)}\bgroup}%
    l%
    <{\egroup}%
}

\usetikzlibrary{patterns}
\usetikzlibrary{arrows}
\pgfdeclarepatternformonly{north east lines wide}{\pgfqpoint{-1pt}{-1pt}}{\pgfqpoint{10pt}{10pt}}{\pgfqpoint{9pt}{9pt}}%
{
  \pgfsetlinewidth{0.4pt}
  \pgfpathmoveto{\pgfqpoint{0pt}{0pt}}
  \pgfpathlineto{\pgfqpoint{9.1pt}{9.1pt}}
  \pgfusepath{stroke}
}

\newcommand{\bi}{\begin{itemize}}
\newcommand{\ei}{\end{itemize}}
\newcommand{\be}{\begin{enumerate}}
\newcommand{\ee}{\end{enumerate}}
\newcommand{\tion}[1]{\S\ref{sect:#1}}
\newcommand{\fig}[1]{Figure~\ref{fig:#1}}
\newcommand{\tab}[1]{Table~\ref{tab:#1}}

\newcommand{\alg}[1]{Algorithm~\ref{alg:#1}}
\newcommand{\algl}[2]{line~\ref{alg:#1:#2}}

\tikzset{%
    body/.style={inner sep=0pt,outer sep=0pt,shape=rectangle,draw,thick,pattern=north east lines wide},
    dimen/.style={<->,>=latex,thin,every rectangle node/.style={fill=white, midway,font=\small}},
    symmetry/.style={dashed,thin},
}

\usepackage[leftmargin=2em,rightmargin=2em]{quoting}

\begin{document}
\title{RIOT: a Stochastic-based Method for Workflow Scheduling in the Cloud}

\author{\IEEEauthorblockN{Jianfeng Chen, Tim Menzies}
\IEEEauthorblockA{Computer Science, North Carolina State University, USA\\
jchen37@ncsu.edu, tim@menzies.us}
}
\maketitle

\begin{abstract}
Cloud computing provides engineers or scientists a  place to run
complex computing tasks. Finding a workflows's  deployment configuration in  a cloud environment is not easy. Traditional workflow scheduling algorithms were based on some heuristics, e.g. reliability greedy, cost greedy, cost-time balancing, etc., or more recently, the meta-heuristic methods, such as genetic algorithms.
These methods are very slow and not suitable for rescheduling in dynamic cloud environment.

This paper introduces RIOT (Randomized  Instance Order Types), a stochastic based method for workflow scheduling.
RIOT groups the tasks in the workflow into virtual machines via a probability model and then uses an effective surrogate based method to assess large amount of potential schedulings.
Experiments in dozens of study cases showed that RIOT 
executes tens of times faster than traditional methods while
generating   comparable results to other methods.
\end{abstract}


\section{Introduction}\label{sect:introduction}

\input{images/structure}

Scientific workflows (a.k.a data-intensive workflow) such as those shown in 
\fig{structure}
have been widely applied 
in scientific research, data mining and business intelligence analysis
~\cite{vockler2011experiences}.
One fundamental problem in workflow research is  {\em workflow scheduling}, i.e. associating the appropriate computer resource  to each task in the workflow.


For  complex workflows, prior work  used meta-heuristic optimizers
(genetic algorithm, particle swarm optimization, ant colony optimization, etc.
\cite{topcuoglu2002performance,rodriguez2017taxonomy,chen2009ant,dorigo2006ant,rodriguez2014deadline,shi2001particle,tsai2014hyper,zhu2016evolutionary}).
Such meta-heuristics are often   computationally expensive. 
For example, in this paper,
we applied a current state-of-the-art meta-heuristic workflow scheduling
algorithm (EMSC~\cite{zhu2016evolutionary}) to find optimized scheduling for 20 workflows. The  the total runtime for optimization was more than 10 hours
(on desktop computer with 2.0GHz, 8GB memory). As a comparison, the total expected runtime for these  workflows was around 9 hours. In some specific workflows, the optimized time is 14x longer than its expected runtime in Amazon AWS clouds.

Such long runtimes are problematic.
Cloud computers execute in highly dynamic environments
where scheduling
tools need to be adaptive to changing conditions
~\cite{rehman2010initial,schad2010runtime,iosup2011performance}.
For example, Schad {\it et al.}~\cite{schad2010runtime} found that the runtime
of a widely used
benchmarks suite can vary by up to 33\% even when run on supposedly identical
instances within the same cloud environment. Not only CPU, but also bandwidth can be highly
variable within the cloud. Schad {\it et al.} report that  network bandwidth between
the same type of EC2 instances  can vary from 410KB/s to 890KB/s.  
Hence,
even after a workflow is planned and deployed, it is important to monitor instances
and repeat the scheduling process during deployment when necessary. 
If   repeated reschedulings are too slow, then it becomes impractical to use
those algorithms.

Accordingly, this paper explores methods for faster scheduling of workflows.
RIOT (short for Randomized Instance Order Types)
is a stochastic  method that does not need not to
evaluate large amount of potential scheduling plans. Instead, it first    applies
some simple heuristics to cluster workflow before using
``surrogate sampling algorithm''
to find proper computing resource for each cluster of tasks.
When  compared  to several existed tools
(including a widely used heuristic method and a state-of-the-art meta-heuristic method), 
RIOT performs the optimization significantly (1-29x) faster than other
methods. Better yet, 
 RIOT's faster analysis found  schedules  as good, or better,
 than  those from other methods
(especially  for  large workflows).

It is important to note that RIOT is a workflow {\em scheduling} algorithm
rather than a workflow {\em prediction} algorithm.  ``Prediction'' refers to determine the expected runtime (workload) of tasks in a workflow. There has been some recent work showing that such predictions are possible. For example, Singh {\it et al.}~\cite{singh2017machine} created a machine learning approach for scientific workflow performance predictions; Hsu {\it et al.}~\cite{hsu2016inside} achieved 91.1\% accuracy on predicting performance of the investigated distributed systems. 

But this paper is not about prediction. Rather, it is about ``scheduling''; i.e. selecting virtual machines, then dividing the workflow across those machines. Due to the complexity of cloud environments, schedulers know that their decisions will only be approximately optimum (and may have to be revised once we learn more about  the running system).

The rest of this paper is organized as follows.
Section~\ref{sect:workflow_scheduling} formulates the workflow scheduling and discusses existed related work in this area.
Section~\ref{sect:algorithm} describes our methodology. 
Section~\ref{sect:experiment}
assess RIOT using dozens of real-world scientific workflows.
Section~\ref{sect:discussion} and~\ref{sect:conclusions} further discuss RIOT.
Note that a reproduction for this work is available on-line at 
 https://github.com/ai-se/riot.

\section{Workflow Scheduling on Cloud Environment}\label{sect:workflow_scheduling}

\subsection{Problem Formulation}\label{sect:definition}
Scientific workflows, a.k.a. data-intensive workflows
typically contain many computational tasks.  These tasks are commonly interconnected via data or resource dependencies.
One common way to represent the dependencies is through Directed Acyclic Graph ({DAG})
as shown in \fig{structure}.
For a DAG $D=\langle V,E\rangle$,  each task is represented as a vertex and every edge $e(i,j)$ indicates that
task $j$ must be executed after  task $i$ is finished.
Mathematically, we denote
\[
\begin{array}{r@{~=~}l}
\mathit{Pred}(i)& \{j|(j,i)\in E\}\\
\mathit{Succ}(i)&\{j|(i,j)\in E\}
\end{array}
\]
Task $i$ can start only after all tasks of $\mathit{Pred}(i)$ are terminated.
For convenience, among all tasks, we denote $T_s$ as the {\it start} task which has no predecessors; and $T_e$, on the other hand, as the {\it exit} task without any successors. 
In this paper, 
we assume that all workflows have single start task and exit task
(this can be simply assured by adding dumb vertex as the head(tail) of all start(exit) tasks).

When deploying workflow into cloud environment, such as Amazon AWS services or Microsoft Azure Cloud,  tasks can be executed in different {virtual machines}. Input/output files of tasks
can be transferred via networking.
Parallel executing in multiple virtual machines significantly reduced execution time of whole workflow, compared to
trivial single PC execution.
A deployment plan for the workflow under the specific cloud environment can be uniquely
determined by following three components (see \fig{workflow_demo} as an example)~\cite{zhu2016evolutionary}:

\begin{figure}[t]
    \centering
    \begin{tikzpicture}[fill=black!20, scale=0.85, transform shape]
    \tikzset{vertex/.style = {shape=circle,draw,minimum size=0.8em}}
    \node[vertex,fill] (ts) at  (0,1.5) {$T_s$};
    \node[vertex] (t1) at  (1,2) {1};
    \node[vertex] (t2) at  (1,1) {2};
    \node[vertex] (t3) at  (2,2) {3};
    \node[vertex] (t4) at (2,1) {4};
    \node[vertex] (t5) at (3,1.5) {5};
    \node[vertex,fill] (te) at (4,1.5) {$T_e$};
    
    \node[text width=2.9cm, fill=black!10] at (0, -0.8) {
    {\bf task2VM mapping } \\
    $T_s,1,2::\mathit{VM}_a$\\
    ~~~~$3,4::\mathit{VM}_b$\\
    ~~$5, T_e::\mathit{VM}_c$
    };
    
    \node[text width=3.5cm, fill=black!10] at (3.6, -0.8) {
    {\bf VM types } \\
    $\mathit{VM}_a, \mathit{VM}_c::$ t2.small\\
    ~~~~~~~~$\mathit{VM}_b::$ t2.large\\
    ~
    };
    
    \node[text width=6.85cm, fill=black!10] at (2, -2.3) {
    {\bf secondary ordering } \\
    $T_s, 1,2,3,4,5,T_e$
    };

    
    

    \draw[->] (ts) to (t1);
    \draw[->] (ts) to (t2);
    \draw[->] (t1) to (t3);
    \draw[->] (t2) to (t4);
    \draw[->] (t3) to (t5);
    \draw[->] (t4) to (t5);
    \draw[->] (t5) to (te);

    \end{tikzpicture}
    \caption{
    A scheduling example that uniquely set the scheduling. 
    In this workflow,
    Task $T_s, 1, 2$ are deployed in  $\mathit{VM}_a$(an AWS EC2 t2.small instance). Task 3, 4 are deployed in $\mathit{VM}_b$ (an AWS EC2 t2.large instance), etc. If both task 3 and 4 are ready to run at one moment,  $\mathit{VM}_a$ will run task 3 first, since task 3 has higher rank.
    }
    \label{fig:workflow_demo}
\end{figure}

\bi
\item {\bf task2VM mapping}: number of VMs should be used and what tasks should be deployed in the same VM.
\item {\bf VM types}: which type of computational resources should be assigned to the VMs.
\item {\bf secondary ordering}: ordering tasks within the same VM. 
Tasks inside one VM are polling in secondary ordering until one of them 
is ready to run.
\ei

In this work, we treat workflow as a multi-objective problem.
The goal of RIOT is to minimize execution time as well and the cost of hiring the
virtual machines from cloud service provider.

For each task $i$, denote
\begin{eqnarray}
\mathit{ft}(i) &=& \mathit{st}(i) + \mathit{dur}(i) \\
\mathit{dur}(i) &=& \mathit{workloads}(i) + \mathit{filetime}(i)\\
\mathit{filetime}(i) &=& \sum_{j\in \mathit{Succ}(i)}^{
\mathit{VM}(i)\neq\mathit{VM}(j)
} \frac{\mathit{file}(i,j)}{\mathit{min}[\mathit{bw}(i), \mathit{bw}(j)]}
\end{eqnarray}
where $\mathit{ft}(i), \mathit{st}(i), \mathit{dur}(i), \mathit{filetime}(i)$ are finish time, start time, duration and I/O
time of task $i$ respectively.
Duration of task $i$ includes computing time of the workload as well as I/O time. $\mathit{file}(i, j)$ is the data flow between task $i$ and task $j$.
I/O speed is limited by  bandwidth($\mathit{bw}$) of VMs.
Then, ``execution time''
is   measured as 
{\it makespan}; i.e. $\mathit{ft}(T_e)$.

As to cost, in scientific workflows,
we create a virtual machine when any task need it and terminate it only when no more future task needs it. 
Hence, the cost of a workflow is the sum of cost of every used virtual machine.
Cost rate and charging policy may differ among  providers
so for this paper,
we followed AWS EC2 pricing (as done by previous works~\cite{zhu2016evolutionary,durillo2014multi,rodriguez2014deadline}).
Note that the charging unit of VM by the hour.

A multi-objective optimizer should return is the {\it Pareto Frontier}.
Mathematically, in this problem, we define one scheduling $s_i$ dominates another scheduling $s_j$ iff

{\small \begin{equation}
\label{eq:domination}
\mathit{makespan}(s_i)\le\mathit{makespan}(s_j)\text{and } \mathit{cost}(s_i)\le\mathit{cost}(s_j)
\end{equation}}
All schedulings not dominated by any others form the Pareto Frontier. Engineers can inspect this frontier to find the solutions they find most useful. 

\subsection{Related Work}\label{sect:relatedwork}

In the literature, there are two ways to address the above problem:
{\em decision-support tools} and {\em automatic optimizers}. 
In these
decision-support tool, the final schedules are determined by humans after reflecting on the information offered by the tool. 
For example tools like  iCanCloud~\cite{nunez2012icancloud} are  
ElasticSim~\cite{cai2017elasticsim} has a GUI platforms where a human
analyst can manually look at the effects of trying a few options.

Automatic optimizers like RIOT are different to decision-support tools in two important ways.
Firstly, the final schedules are determined by automatic algorithms.
Secondly, these automatic optimizers explore many more options that humans
could ever manage within a decision-support tool.

Many   researchers have explored a wide range of  automatic optimization methods to find
cloud computing scheudles.   One of the earliest was
Topcuoglu {\it et al.}~\cite{topcuoglu2002performance}, who proposed  a heuristic method,
 HEFT (Heterogeneous-Earliest-Finish-Time). HEFT has two phases, task prioritizing phase and processor (VM) selection phase.
In task prioritizing phase, tasks were ranked by their computation as well as communication cost.
In processor selection phase, tasks were assigned to the processor which was first available.


Latter as  workflows become  larger and larger, experiment showed that HEFT's heuristics were easily trapped into local optimal~\cite{rodriguez2017taxonomy}. Therefore, 
researchers turned
to   meta-heuristic methods. 
Chen {\it et al.}~\cite{chen2009ant} used  Ant Colony Optimization (ACO)~\cite{dorigo2006ant}.
In that work,
seven heuristics were applied
to propose a  pheromone function, such as reliability greedy, cost greedy, time/cost balance etc. 
Rodriguez {\it et al.}~\cite{rodriguez2014deadline}
found that particle
swarm optimization (PSO)~\cite{shi2001particle} outperformed ACO as well as much other prior work.
At the same time, 2014, Tsai {\it et al.}~\cite{tsai2014hyper} proposed  {HHSA} (hybrid heuristic-based scheduling algorithm) framework. HHSA was an ensemble method that ran separate
ACO, PSO, and other meta-heuristic algorithms,  then reported the best solution
found by any method.

\begin{figure}[!t]
\small
~\hrule~
\begin{enumerate}[label=\alph*, leftmargin=10pt]
    \item Generate population $i=0$ using some {\em initialization policy}.
    \item Evaluate all individuals in population 0.
    \item Repeat until tired or happy
        \begin{enumerate}[label=c.\arabic*]
        \item {\em Cross-over:} combine elite items to make population $i+1$;
        \item {\em Mutation:} make small changes within population $i$;
        \item {\em Evaluate:} individuals in population $i$;
        \item {\em Selection:} choose some elite subset of population $i$.
        \end{enumerate}
\end{enumerate}

~\hrule~
\caption{Framework of Multi-objective Evolutionary Algorithms
such as NSGA-II, SPEA2, MOEA/D.
}
\label{fig:moea}
\end{figure}
\fig{moea} shows the general framework for one type of meta-heuristic methods -- multi-objectives evolutionary algorithms (MOEA).
Variants of the MOEA, including SPEA2, NSGA-II and  
MOEA/D etc., differ in how they down-select from the general
population to the frontier as defined in \S\ref{sect:definition}.
Workflow researchers applied various evolutionary algorithms to solve 
the scheduling problem with multi-objectives. 
Among them, our reading of this literature
is that the Zhu {\it et al.}~\cite{zhu2016evolutionary} paper on EMSC is a comparative
assessment of much of the previous work.
EMSC combined three scheduling components (as defined in \S\ref{sect:definition}) into 
an integer array and created unique reproduction (cross-over and mutation) operators.
Zhu {\it et al.} showed
that   off-the-shelf evolutionary algorithms such as NSGA-II, SPEA2, MOEA/D could be effectively 
figure out optimal schedulings.
The evaluation part of the Zhu {\it et al.} paper is
every extensive and showed that EMSC
achieved better schedulings that a wide
range of other approaches. 

Another recent work came from Wang {\it et al.}~\cite{wang2017using}.
Wang {\it et al.} formulated the workflow scheduling problem using 
mixed integer programming.
MIP was dedicated to to find the {\it global} optimal cost under some deadline constraints.
However, the runtime for MIP algorithm was quite long-- they needed hours to figure out solutions for workflows with hundreds of tasks, which was similar to Zhu's EMSC framework. Therefore, their method are recommended for scheduling small or medium workflows.
RIOT, on the other hand, is able to find the near-optimal scheduling for large workflows in minutes.




One frequently asked question we get about RIOT is 
``why not set up a deadline requirement and try to optimize the single objective--monetary cost?''. Solving single objective problems is certainly much easier than solving multi-objective problems. 
However, prior results~\cite{zhu2016evolutionary} showed that in many times, by relaxing the deadline requirement a little bit (extend by 5\%), the monetary cost may significantly reduce (by more than 30\%).

\section{How to Make a  RIOT} \label{sect:algorithm}
\subsection{Overview}

The design of RIOT was motivated by the following key observation: all
the above methods  {\em  configured the scheduling at the task level}:
\bi
\item
The heuristic HEPT algorithm assigned the tasks one by one (each assignment need a traversal of all VMs); 
\item
The meta-heuristic  EMSC algorithm  applied an integer bit  to encode every task and performed the evolution basing on that encoding.
\ei
RIOT was designed to test the following conjecture:  {\em better
schedules can be found, faster, if we schedule at  a higher level;} i.e. using group of tasks in the VMs.
Hence, as described in this section:
\bi
\item
RIOT's {\sc B-Rank} orders tasks by prior knowledge and structure of the workflow.
\item
RIOT's {\sc TaskGroup} sub-routine uses the   concept of {\it critical task} (described below)
to divide large amount of tasks into much fewer blocks. 
\ei
Another key conjecture of RIOT is  the {\em anchor hypothesis}.
That is, when evaluating candidates, {\em it is sufficient to extrapolate between just a few
anchor points}:
\bi
\item
RIOT   evaluates a very small number of randomly selected candidates, which we call the {\em anchors}.
For the anchor evaluation, RIOT uses  the CloudSim workflow simulator~\cite{calheiros2011cloudsim}.
This is a very widely-used approach-- as of Feb 2018, Google scholar reports
that the original 2011 CloudSim paper has 2,500+ citations.
\item
RIOT evaluates the other candidates by extrapolating between the  anchors (and we call this
extrapolation {\sc SurrogateEvaluate}s). 
\ei
Just to say the obvious:  anchor evaluation is much faster than evaluating
all candidates since only the anchors require time-consuming simulations or executions on real hardware.



Algorithm~1 shows the details of  RIOT. The rest of this section discusses those details.

\input{images/framework_alg}

\input{images/grouping}

\subsection{Components of RIOT}\label{sect:components}
\subsubsection{\colorbox{black!15}{\scshape B-Rank}}\label{sect:brank} 

RIOT used the B-Rank method of HEFT~\cite{topcuoglu2002performance} to sort task execution priorities on the virtual machine.
When sorting tasks, the B-rank metric is distance of the activity to the end of the workflow.
Let $T_e$ be  the final task of a workflow (defined in~\tion{definition}). Then 
\[
\begin{array}{r@{~=~}l}
\mathit{rank}(T_e) & 1 \\
\mathit{rank}(i)  & 1+\max\limits_{j \in \mathit{Succ}(i)} \mathit{rank}(j)  \end{array}
\]

B-rank sorts tasks in  decreasing order of $\mathit{rank}(i)$.
Then, if two 
tasks $i,j$ are ready to run on the same virtual machine,
we select the task that is {\em lower} in this sort order
(and ties are resolved randomly).

\subsubsection{\colorbox{black!15}{\scshape TaskGroup}}\label{sect:grouptasks}
To map tasks into virtual machines, RIOT use {\sc TaskGroup} to cluster the tasks, and assign each cluster into one VM.

{\sc TaskGroup} defines the {\it critical tasks} as follows,
  \begin{itemize}
  \item It is the start task $T_s$, or
  \item  It is a 
   task whose data flow in-degree (number of edges incident to in workflow) is among the top third of all in-degress of all tasks
   \end{itemize} 

Next {\sc TaskGroup} assigns each task a probability $p_i$ as 
\begin{equation*}
    p_i = \begin{cases}
               1.0               & \mathit{task}_i  \text{is critical task}\\
               \eta * \underset{j \in \mathit{Pred}(i)}{\mathit{average}}(p_j) & \text{otherwise}
           \end{cases}
\end{equation*}
where $\eta\in[0, 1]$ is a control parameter.

With the probabilities, we can group tasks into the clusters.
For each task, there is $p$ probability to assign a new cluster. If one task does not map to new cluster,
{\sc TaskGroup} map it to any one of existed clusters (if it is a
critical task) or any clusters of its predecessors (if it is not a critical task).

\fig{grouping} demonstrates critical tasks and $p$ assignment.
As we can see, critical task separates the workflow into several ``blocks''. Tasks within one block are supposed to be executed in serial, therefore, in the same cluster.
$p$ value can control this-- tasks closer to end of workflow have smaller $p$ within a block, therefore, higher probability to assign to its predecessor's cluster.

From \fig{grouping}, we can see that $\eta$ is an parameter controlling
number of clusters.
Higher $\eta$ implies more virtual machines might be applied in the deployment.
To improve diversity (and explore more solutions), we set $\eta$ as 0.05, 0.10, $\ldots$ , 1.00 to generate 20 different task-instance mappings.




\begin{table}[!t]
\caption{Eight types of available EC2 virtual machines (On-Demand instances in US East N.Virginia), sorted by price. 
}
\centering
\small
\begin{tabular}{rccc}
    \toprule
    Type & Compute Unit & Bandwidth (MB/s) & Price(\$/hr)\\
    \midrule
  m3.medium & 3.75 & 85.2 & 0.067 \\
  m4.large & 7.5 & 35.2 & 0.1 \\
  m3.large & 7.5 & 85.2 & 0.133\\
  m4.xlarge&15 &  68& 0.2 \\
  m3.xlarge& 15 & 131 & 0.266\\
  m4.2xlarge&30 & 131& 0.4\\
  m3.2xlarge&40 & 131& 0.532\\
  m4.4xlarge&45 & 181&0.8\\
    \bottomrule
  \end{tabular}
\label{tab:iaasParameter}
\end{table}

\subsubsection{\colorbox{black!15}{\scshape SurrogateEvaluate}}\label{sect:sample}


Nowadays most cloud service provider provide many predefined virtual machines 
such as the EC2 instances of Table~\ref{tab:iaasParameter}. These machines  differ in computational abilities, bandwidth, and certainly, the unit rental price.
After grouping the tasks into VMs, we should find proper VM-type mappings (short for {\it mapping(s)} in this section).
One naive assumption was that the function $f:\mathit{mappings}\rightarrow\mathit{objective(make/cost)}$ is convex. If so, we can simply applied hill climbing (HC)\footnote{For space reasons, HC and latter mentioned SA experiments are not shown here. See \url{http://tiny.cc/qptcry}.}. Unfortunately, with HC, results were worst than existed methods. That is, $f$ is not convex. 

Hajela~\cite{hajela1990genetic} indicated that SA can better handle non-convex problems. 
Results showed that SA did perform better than HC. This approved ``$f$ is not convex''.
However, SA still under-performs existed method. Having SA results, we supposed that we should randomly explore LARGE amount of mappings and select the best ones. 

{\sc SurrogateEvalaute} try to create large number of mappings, and assess them so that we can pick up the best mapping for each {\it Task2VM} clustering setting.
\alg{sample} illustrates this process.

RIOT scores candidates by extrapolating between a small number of {\em anchors} (i.e.
a very small number of mappings with  known  evaluations  cores  (line  4-12),  including $n_0$ random mappings plus $n_T$ iso-mapped instances. (i.e. uniformed type of VMs). 
Next we use surrogate-based method to assess other large amount of
random mappings.
For each vm-type mapping to be assessed, we find the nearest $a_n$ and furthest $a_f$ mappings in evaluated anchors; and then guess its $o$ (makespan or cost) through following formula (equivalent to \algl{sample}{ge})
\begin{equation}
\label{eq:guess}
\frac{\widehat{o}_r- o_n}{o_f - o_n} = \frac{\mathit{dist}(a_n, r) }{\mathit{dist}(a_n, a_f)}cos\theta
\end{equation}
where $o_f, o_n, \widehat{o}_r$ are objectives of $a_f$, $a_n$ and expected objective of $r$ respectively; $\mathit{dist}(\cdot, \cdot)$  is the distance (defined next) between two
vm-type mappings.

Typically the vm with higher unit rental price is equipped with better CPU, memory and bandwidth. Consequently, we can sort the {\it TYPES} and assign each a rank.
With this, distance between two vm-type mappings $X,Y$ is defined as
\begin{equation}
\mathit{dist}(X, Y) = \left(\sum (x_i-y_i)^\alpha\right)^{1/\alpha}
\end{equation}
$x_i(y_i)$ is the ranking of virtual machine used for $i$-th cluster according to $X(Y)$. 
Most machine learning algorithm set $\alpha=2$ (the Euclidean distance).
In RIOT, we found that using $\alpha=1$ (the Manhattan distance) returned similar results as Euclidean distance.
As Aggarwal {\it et al.}~\cite{aggarwal2001surprising} argues, Manhattan metric is preferred to the standard Euclidean metric when data dimension is large.

\input{images/sample_alg}

 Algorithm~\ref{alg:sample} sumamrizes the 
{\sc SurrogateEvalaute} algorithm discussed in this section.
\section{Evaluation}\label{sect:experiment}

In this section we report numerical results of scheduling workflows in different structures as shown in \fig{structure}. Each structure scales 25  to 1000 tasks approximately (see \url{http://tiny.cc/wfeg} for more details).
All workflows were supposed to deploy to {Amazon AWS Cloud Services}, with instances listed in \tab{iaasParameter}.

In this paper, we compared RIOT to two baseline schedulers. First is MOHEFT~\cite{durillo2014multi} (Multi-objective HEFT), a {\bf heuristic algorithm} basing on the classic HEFT~\cite{topcuoglu2002performance} method.
Another is EMSC~\cite{zhu2016evolutionary}, a {\bf meta-heuristic algorithm}.
We ran EMSC basing on three popular MOEA, including NSGA-II, SPEA2 and MOEA/D.

We coded RIOT and two baseline tools in JAVA and ran them one the same machine (2.0GHz with 8GB memory, running in CentOS).
For parameters of RIOT, by default, we set
$N,n_0,n_T=\{500, 30, 8\}$
since we are using the eight types of Table~\ref{tab:iaasParameter}.
For other baselines, we strictly 
follow the setups defined in their associated publications.

To test  performance robustness and  reduce observational error,
we repeated these all studies 30 times with different random seeds.
To check the statistical significance of the differences between the algorithms, 
we performed a statistical test using Wilcoxon test at a 5\% significance level.

\subsection{Comparing via Runtime}\label{sect:runtime}

\input{tables/t_runtime}
\input{tables/bigtable}
Table~\ref{tab:runtime} shows the runtimes
of  different treatments. 
For convenience, we also report the {\it makespan} as well as speed-up of RIOT method.
From this table we observe:
\bi
\item
Measured in relative terms, except in a very small workflow (Montage25), we note that RIOT
is 1-27x faster than other approaches.
\item
Measured in absolute values, the general trend is that other methods can take up to 4.8 hours while RIOT is never slower than 310 seconds. That is to say, RIOT terminates in 
just a few seconds to  minutes  
while other methods require minutes to hours. 
\item Comparing the expected makespan to optimizer runtimes, in some workflows with many tasks, such as Montage100, Montage1000, CyberShake1000, etc., previous methods were slower than the
eventual runtimes (makespan); while RIOT requires just small ratio of makespans, making RIOT more suitable for re-scheduling in dynamic cloud environment.
\ei

To conclude, comparing via runtime,
{\bf RIOT finds schedulings much faster than the prior heuristic/meta-heuristic methods.}

\subsection{Comparing via Frontier Quality}\label{sect:quality}
In this paper we treat workflow scheduling as a multi-objective problem. 
To compare the  quality of returned frontier, three measures is widely applied~\cite{marler2004survey} -- Hypervolume, Inverted Generation Distance (IGD) and Spread. For our problem, we plot the objectives of frontiers in a 2D coordinate
and have following definitions, 
\bi
\item {\it Hypervolume} is the area of space the obtained frontier dominated (top-right of the frontier curve);
\item {\it IGD} is average Euclidean distance of each point in obtained frontier to its nearest point in the true Pareto Frontier. It is almost impossible to find the true Pareto Frontier. Following  Wang {\it et al.}'s guidance~\cite{wang2016practical}, we collected non-dominated scheduling found by any algorithms in any repeats as the true frontier;
\item {\it Spread} defines the average Euclidean distance of every pair of consecutive points in the obtained frontier. Lower spread implies better diversity.
\ei



\tab{whole} concludes the statistical measures  for all workflows.
In that table, {\bf Bold}
values indicate where RIOT performed as well as or better than any of MH/EN/ES/EM (under the Wilcoxon Test).

Within \tab{whole}, we observe that:
\bi
\item Variants of EMSC have similar performance; EMSC outperforms the heuristic method, MOHEFT. This is consisted with EMSC's origin paper;
\item Consider the IQR values\footnote{IQR = intra-quartile range =  (75-25)th percentile.}: the performance of RIOT is stable, even though it is a stochastic method;
\item Measured by hypervolume, in 70\% of experimented workflows, RIOT has significantly higher values than other baselines.
That is, in most workflows, with monetary cost constraint, RIOT can find schedulings with less makespan, or within some deadline, RIOT can find schedulings required less cost;
\item Measured by IGD, RIOT performs best in 85\% workflow. In other words, RIOT's results are closer to true frontier;
\item According to spread statistics, RIOT provides more diverse results in majority study cases;
\item Summarizing all in \tab{whole}, there is no any algorithm always performs the best (this is one of Wolpert's {\it No Free Lunch} results~\cite{Wolpert97}). However, RIOT performs best in majority of measurements
and no bad in the remaining measures (compared to {\it RAND} or {\it MOHEFT} especially).
\ei

\noindent
Summarizing above observations, {\bf RIOT usually finds schedulings as good as anything else. This result is particularly remarkable for the large workflows}.

\section{Threats to Validity}\label{sect:discussion}

 \subsection{Sampling Bias}
While we tested RIOT on over two dozen workflows,
it would be inappropriate to say that this sample covers the space
of all possible workflows.  As researchers, all we can do is to introduce our method, release the
source code for our method and suggest that other researchers try a broader range of workflows.

\subsection{Algorithm Bias}
In this paper we compared our work to MOHEFT and EMSC. 
We choose MOHEFT, the multi-objective variant of HEFT, since it is one of most popular heuristic method in
this area. It is simple and fast (compared to meta-heuristic methods).
We choose EMSC since it is the best meta-heuristic we have explored in this area. 
As shown in \tion{relatedwork}, there are many other ways
to perform workflow scheduling. For example, the MIP~\cite{wang2017using} can guarantee to get best schedule under some deadline and other constraints. 
Comparing to other methods is left for future work.


\subsection{Evaluation Bias}
Comparing via results  \tab{whole}, we conclude that RIOT is a smart tool.
Note that in some workflows, RIOT did not outperform EMSC  (see three of the Sipht results  of \tab{whole}). 
However,
considering runtimes of different algorithms,
EMSC might requires longer time than expected makespan.
RIOT is significantly faster than other methods.
Therefore, we consider RIOT as a promising tool.
In future, we need to analyze why RIOT fails in some workflows.

\section{Conclusions}\label{sect:conclusions}
This paper introduces RIOT, a  novel stochastic method for workflow scheduling in the cloud.

RIOT was built to test two conjectures: 
\begin{enumerate}
\item
Better
schedules can be found, faster, if we schedule at  a higher level;
\item
When scoring candidates
 it is sufficient
to extrapolate between just a few anchor points.
\end{enumerate}
The results shown above strongly endorse these two conjectures.
When optimizing large workflows, experiments showed that about 80\% of RIOT's quality indicators were as good or better than existed algorithms (MOHEFT and EMSC), but only require less than (1/29=) 3\% of their
optimization time.
As for small or medium workflows, two thirds of
RIOT's quality indicators were as good as other methods, with RIOT  1x-30x faster.

Consequently, we  recommend RIOT for configuring large workflows, since RIOT takes
minutes to find schedulings that other tools need hours to find.
As for smaller workflows, we still recommend researchers try RIOT first since RIOT is much faster than prior work and in the usual case,   RIOT can achieve competitive results
to other methods.

\balance
\bibliographystyle{IEEEtran}
\input{main.bbl}

\end{document}

%% file: images/structure.tex
\begin{figure}[!b]
    \centering
    \noindent
    \begin{tabular}{cccccc}
        \resizebox{!}{60pt}{
        \begin{tikzpicture}
[lineDecorate/.style={-latex,line width=0.2mm}, nodeDecorate/.style={shape=circle,inner sep=3pt,draw}]

 \foreach \nodename/\x/\y in {
0/0/0,
1/1/0,
2/2/0,
3/3/0,
4/1.5/-0.5,
5/1.5/-1.0,
6/1.5/-1.5,
7/1.5/-2,
8/-1/2,
9/0.8/2,
10/2.3/2,
11/3.7/2,
12/-1/1.5,
13/0/1.5,
14/1.2/1.5,
15/2.1/1.5,
16/3/1.5,
17/3.9/1.5,
18/1.5/1.0,
19/1.5/0.5
 }
 {
          \node (T\nodename) at (\x,\y*1.5) [nodeDecorate, fill=black] {};
 }
		
\path
\foreach \startnode/\endnode in {0/4,1/4,2/4,3/4,4/5,5/6,6/7,
8/12,8/13,9/12,9/13,9/14,9/15,10/14,10/16,11/15,11/16,11/17,
12/18,13/18,14/18,15/18,16/18,17/18,
18/19,
19/0,19/1,19/2,19/3}
        {
          (T\startnode) edge[lineDecorate] node {} (T\endnode)
        }

(T8) edge[lineDecorate, bend right=50] node{} (T0)
(T10) edge[lineDecorate, bend left=30] node{} (T2)
(T9) edge[lineDecorate, bend right=30] node{} (T1)
(T11) edge[lineDecorate, bend left=70] node{} (T3);
        \end{tikzpicture}}
         
& 
\resizebox{!}{60pt}{
\begin{tikzpicture}
[lineDecorate/.style={-latex,line width=0.2mm}, nodeDecorate/.style={shape=circle,inner sep=3pt,draw}]

 \foreach \nodename/\x/\y in {
0/1.5/0,
1/1.5/-0.5,
2/1.5/-1
 }
 {
          \node (T\nodename) at (\x*0.8,\y*1.5) [nodeDecorate, fill=black] {};
 }

		 \foreach \x in {0,1,2,3}
 {
          \node (A\x) at (\x*0.8, 4.0) [nodeDecorate, fill=black] {};
		  \node (B\x) at (\x*0.8, 3.0) [nodeDecorate, fill=black] {};
		  \node (C\x) at (\x*0.8, 2.0) [nodeDecorate, fill=black] {};
		  \node (D\x) at (\x*0.8, 1.0) [nodeDecorate, fill=black] {};
 }
 
 \node(H) at (1.5*0.8, 5) [nodeDecorate, fill=black]{};
 
\path
\foreach \startnode/\endnode in {0/1,1/2}
        {
          (T\startnode) edge[lineDecorate] node {} (T\endnode)
        }
\foreach \startnode/\endnode in {H/A1,H/A2,H/A3,H/A0}
        {
          (\startnode) edge[lineDecorate] node {} (\endnode)
 }
 
 \foreach \i in {0,1,2,3}
        {
          (A\i) edge[lineDecorate] node {} (B\i)
		  (B\i) edge[lineDecorate] node {} (C\i)
		  (C\i) edge[lineDecorate] node {} (D\i)
		  (D\i) edge[lineDecorate] node {} (T0)
 };

\end{tikzpicture}}

~~~~~~~\resizebox{!}{60pt}{
\begin{tikzpicture}
[lineDecorate/.style={-latex,line width=0.2mm}, nodeDecorate/.style={shape=circle,inner sep=3pt,draw}]

 \foreach \i in {0,1,2,3,4,5,6,7,8}
 {
          \node (A\i) at (\i*0.6,5*0.8) [nodeDecorate, fill=black] {};
		  \node (B\i) at (\i*0.6,4*0.8) [nodeDecorate, fill=black] {};
		  \node (E\i) at (\i*0.6,2*0.8) [nodeDecorate, fill=black] {};
		  \node (F\i) at (\i*0.6,1*0.8) [nodeDecorate, fill=black] {};
 }
 \node(C) at (2*0.6,3*0.8) [nodeDecorate, fill=black]{};
 \node(D) at (6.5*0.6,3*0.8) [nodeDecorate, fill=black]{};
 \node(G) at (2*0.6,0*0.8) [nodeDecorate, fill=black]{};
 \node(H) at (6.5*0.6,0*0.8) [nodeDecorate, fill=black]{};
  \node(I) at (4.5*0.6,-1*0.8) [nodeDecorate, fill=black]{};
\path

 \foreach \i in {0,1,2,3,4,5,6,7,8}
        {
          (A\i) edge[lineDecorate] node {} (B\i)
		  (E\i) edge[lineDecorate] node {} (F\i)
 }
  \foreach \i in {0,1,2,3,4}
        {
          (B\i) edge[lineDecorate] node {} (C)
		  (C) edge[lineDecorate] node {} (E\i)
		  (F\i) edge[lineDecorate] node {} (G)
 }
  \foreach \i in {5,6,7,8}
        {
          (B\i) edge[lineDecorate] node {} (D)
		  (D) edge[lineDecorate] node {} (E\i)
		  (F\i) edge[lineDecorate] node {} (H)
 }

 (G) edge[lineDecorate] node {} (I)
 (H) edge[lineDecorate] node {} (I);
\end{tikzpicture}}

\\

\resizebox{!}{40pt}{
\begin{tikzpicture}
[lineDecorate/.style={-latex,line width=0.2mm}, nodeDecorate/.style={shape=circle,inner sep=3pt,draw}]

 \foreach \i in {0,1,2,3}
 {
          \node (A\i) at (\i*0.6+0.15,1*0.8) [nodeDecorate, fill=black] {};
		  \node (B\i) at (\i*0.6,0*0.8) [nodeDecorate, fill=black] {};
		  
 }
 
  \foreach \i in {5,6,7,8}
 {
          \node (A\i) at (\i*0.6-0.15,1*0.8) [nodeDecorate, fill=black] {};
		  \node (B\i) at (\i*0.6,0*0.8) [nodeDecorate, fill=black] {};
		  
 }
 
 \node(X) at (1.5*0.6, 2*0.8) [nodeDecorate, fill=black]{};
 \node(Y) at (6.5*0.6, 2*0.8) [nodeDecorate, fill=black]{};
\node(K) at (4*0.6, 0*0.8) [nodeDecorate, fill=black]{};
\node(E) at (4*0.6,-1*0.8) [nodeDecorate, fill=black]{};
\path

 \foreach \i in {0,1,2,3}
        {
          (A\i) edge[lineDecorate] node {} (B\i)
		  (A\i) edge[lineDecorate] node {} (K)
		  (B\i) edge[lineDecorate] node {} (E)
		  (X) edge[lineDecorate] node {} (A\i)
 }
 (K) edge[lineDecorate] node{}(E)
  \foreach \i in {5,6,7,8}
        {
          (A\i) edge[lineDecorate] node {} (B\i)
		  (A\i) edge[lineDecorate] node {} (K)
		  (Y) edge[lineDecorate] node {} (A\i)
		  (B\i) edge[lineDecorate] node {} (E)
 };
 
\end{tikzpicture}}

&

\resizebox{!}{40pt}{
\begin{tikzpicture}
[lineDecorate/.style={-latex,line width=0.2mm}, nodeDecorate/.style={shape=circle,inner sep=3pt,draw}]

 \foreach \i in {0,1,2,3,4}
 {
          \node (A\i) at (\i*0.6,-1*0.8) [nodeDecorate, fill=black] {};
		  
 }
 
  \foreach \i in {0,1,2,3}
 {
          \node (C\i) at (\i*0.6+0.2,1*0.8) [nodeDecorate, fill=black] {};
		  
 }
 
   \foreach \i in {0,1,2,3,4,5,5,6,7,8,9,10}
 {
          \node (F\i) at (\i*0.4+1.8,0.2*0.8) [nodeDecorate, fill=black] {};
		  
 }
 
 \node(B) at (1.5*0.6, 0*0.8) [nodeDecorate, fill=black]{};
 \node(D) at (2.8*0.6, -2*0.8) [nodeDecorate, fill=black]{};
  \node(E) at (5*0.6, -1*0.8) [nodeDecorate, fill=black]{};
\path
(E) edge[lineDecorate] node {} (D)  
 \foreach \i in {0,1,2,3}
        {
          (C\i) edge[lineDecorate] node {} (B)  
 }
  \foreach \i in {0,1,2,3,4}
        {
          (A\i) edge[lineDecorate] node {} (D)
		  (B) edge[lineDecorate] node {} (A\i)
 }
   \foreach \i in {0,1,2,3,4,5,5,6,7,8,9,10}
        {
          (F\i) edge[lineDecorate] node {} (E)
 };
\end{tikzpicture}}

\\
    \end{tabular}
    \caption{
  Examples of cloud computing  workflows. Clockwise from top left: {\tt Montage, Epigenomics, Inspiral, CyberShake, Sipht}. Each node is one ``task''
    and each edge is a data flow from one task to another.   Number of tasks  can vary from dozens to   thousands).  The scheduling problem is to map these tasks to a smaller number
    of virtual machines, decide the ordering of tasks within one VM, and
    then decide what kind of machine should drive each VM. 
    }
    \label{fig:structure}
\end{figure}

%% file: images/framework_alg.tex
\begin{algorithm}[!t]
\footnotesize

\SetKwInOut{Input}{Input}
\SetKwInOut{Output}{Output}
\SetKwInOut{Parameter}{Parameter}
\SetKwRepeat{Do}{do}{while}
\Input{{\it DAG}, directed acyclic graph representing the workflow\\
{\it TYPES}, set of available virtual machine types\\ 
{\it CloudSim}, simulator}
\Output{$S_B$, Best schedulings with balanced makespan and cost}
\BlankLine
\begin{spacing}{1.15}
\BlankLine
Order $\leftarrow$ {\scshape B-Rank}({\it DAG})\label{alg:framework:order}\;
$S$ $\leftarrow \emptyset$\;
\ForEach{$\eta \in [0.05,0.1,\ldots,1.0]$}{
Task2VM
$\leftarrow${\sc TaskGroup}({\it DAG}, $\eta$)\label{alg:framework:group}\;
$\{\mathcal{T}\}$ $\leftarrow$ {\sc SurrogateEvaluate}(Task2VM, {\it TYPES}, {\it CloudSim})\label{alg:framework:surrogate}\;
\ForEach{VMTypes $\in \{\mathcal{T}\}$}{
Add {\it (Task2VM, VMTypes, Order)} to $S$\label{alg:framework:save}\;
}
}

\BlankLine
$S_B \leftarrow$ All non-dominated solutions within $S$\label{alg:framework:a}\;
Evaluate schedulings in $S_B$ by {\it CloudSim}\;
\Return{$S_B$}\label{alg:framework:b}\;
\caption{Framework of RIOT}\label{alg:framework}
\end{spacing}
\end{algorithm}

%% file: images/grouping.tex
\begin{figure}[!t]
    \centering
    \noindent
    \begin{tabular}{cccc}
        \resizebox{!}{90pt}{
        
        \begin{tikzpicture}
    [lineDecorate/.style={->},%
      nodeDecorate/.style={shape=circle,inner sep=3pt,draw}]
    \foreach \nodename/\x/\y/\direction/\navigate in {T_s/1.0/3.5/left/west, a_1/0.5/3.0/left/west, a_2/0.5/2.5/left/west, a_3/0.5/2.0/left/west, b_1/1.0/3.0/below/mid west, b_2/1.0/2.5/below/west,  b_3/1.0/2.0/below/west, c_1/1.5/3.0/right/east, c_2/1.5/2.5/right/east, c_3/1.5/2.0/right/east, d/1.0/1.5/left/west, T_e/1.0/1.0/left/west}
    {
      \node (\nodename) at (\x,\y*1.1) [nodeDecorate] {};
      \node [\direction] at (\nodename.\navigate) {\footnotesize$\nodename$};
    }
    \node [circle, draw, inner sep=3pt, line width=0.55mm] at (1.0,3.5*1.1) {};
	\node [circle, draw, inner sep=3pt, line width=0.55mm] at (1.0,1.5*1.1) {};
    \path
    \foreach \startnode/\endnode in {T_s/b_1, a_1/a_2,a_2/a_3,b_1/b_2,b_2/b_3,c_1/c_2,c_2/c_3,b_3/d,d/T_e}
    {
      (\startnode) edge[lineDecorate] node {} (\endnode)
    }
	\foreach \startnode/\endnode/\dir in {T_s/a_1/right, T_s/c_1/left,a_3/d/right, c_3/d/left}
	{
	 (\startnode) edge[lineDecorate] node {} (\endnode)
	};
\end{tikzpicture}

        } & 
        \resizebox{!}{90pt}{
        
        \begin{tikzpicture}
    [lineDecorate/.style={->},%
          nodeDecorate/.style={shape=circle,inner sep=3pt,draw}]
        \foreach \nodename/\x/\y/\pv in {T_s/1.0/3.5/100, a_1/0.5/3.0/30, a_2/0.5/2.5/9, a_3/0.5/2.0/3, b_1/1.0/3.0/30, b_2/1.0/2.5/9,  b_3/1.0/2.0/3, c_1/1.5/3.0/30, c_2/1.5/2.5/9, c_3/1.5/2.0/3, d/1.0/1.5/100, T_e/1.0/1.0/30}
        {
          \node (\nodename) at (\x,\y*1.1) [nodeDecorate,fill=black!\pv] {};
        }

        \path
        \foreach \startnode/\endnode in {T_s/b_1, a_1/a_2,a_2/a_3,b_1/b_2,b_2/b_3,c_1/c_2,c_2/c_3,b_3/d,d/T_e}
        {
          (\startnode) edge[lineDecorate] node {} (\endnode)
        }
		\foreach \startnode/\endnode/\dir in {T_s/a_1/right, T_s/c_1/left,a_3/d/right, c_3/d/left}
		{
		 (\startnode) edge[lineDecorate] node {} (\endnode)
		};
		\node[text width=1.2cm] at (1.25, 4.3) {\small$\eta=0.3$};
\end{tikzpicture}
        
        } &
        \resizebox{!}{90pt}{
        \begin{tikzpicture}
    [lineDecorate/.style={->},%
          nodeDecorate/.style={shape=circle,inner sep=3pt,draw}]
        \foreach \nodename/\x/\y/\pv in {T_s/1.0/3.5/100, a_1/0.5/3.0/70, a_2/0.5/2.5/49, a_3/0.5/2.0/35, b_1/1.0/3.0/70, b_2/1.0/2.5/49,  b_3/1.0/2.0/35, c_1/1.5/3.0/70, c_2/1.5/2.5/49, c_3/1.5/2.0/35, d/1.0/1.5/100, T_e/1.0/1.0/70}
        {
          \node (\nodename) at (\x,\y*1.1) [nodeDecorate,fill=black!\pv] {};
        }

        \path
        \foreach \startnode/\endnode in {T_s/b_1, a_1/a_2,a_2/a_3,b_1/b_2,b_2/b_3,c_1/c_2,c_2/c_3,b_3/d,d/T_e}
        {
          (\startnode) edge[lineDecorate] node {} (\endnode)
        }
		\foreach \startnode/\endnode/\dir in {T_s/a_1/right, T_s/c_1/left,a_3/d/right, c_3/d/left}
		{
		 (\startnode) edge[lineDecorate] node {} (\endnode)
		};
		\node[text width=1.2cm] at (1.25, 4.3) {\small$\eta=0.7$};

\end{tikzpicture}
        
        } &
        \resizebox{!}{90pt}{\begin{tikzpicture}
\foreach \i in {0,2, 4,...,96}
    \fill[black!\i] (0, \i mm) rectangle ++(5mm,5mm);
\node[text width=1mm] at (10mm,100mm) {\Huge $p$};
\node[text width=5mm] at (-8mm,7mm) {\Huge 0.0};
\node[text width=5mm] at (-8mm,100mm) {\Huge 1.0};
\draw[line width=1pt] (5mm,5mm) -- (-2mm, 5mm);
\draw[line width=1pt] (5mm, 101mm) -- (-2mm, 101mm);
\end{tikzpicture}
}
    \end{tabular}
    \caption{Left: a demonstrated workflow with  critical tasks highlighted;
    Middle\& Right: probability for each tasks to deploy into new VM with different $\eta$;
    $p=0.0$ indicates that task always use existed VM, while $p=1.0$ indicates that
    tasks always request new VMs.
    }
    \label{fig:grouping}
\end{figure}
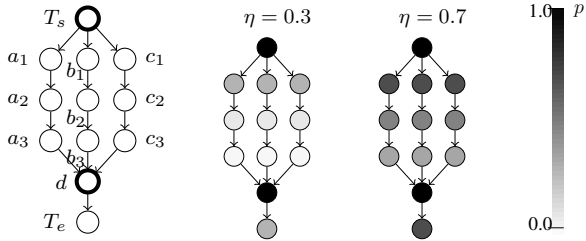

%% file: images/sample_alg.tex
\begin{algorithm}[!t]
\footnotesize
\SetKwInOut{Input}{Input}
\SetKwInOut{Output}{Output}
\SetKwInOut{Parameter}{Parameter}
\SetKwRepeat{Do}{do}{while}
\Input{{\it TYPES}, set of available virtual machine types\\
{\it CloudSim}, simulator}
\Output{$\mathcal{T}$, assessed vm-type mappings}
\Parameter{$N$, number of random vm-type mappings\\
            $n_0$, additional anchor size $(n_0\ll N)$}
\BlankLine

\BlankLine
$n_T \leftarrow$ number of {\it TYPES}\label{alg:sample:a1}\; 
{\it Anchors} $\leftarrow n_T$ iso-mappings $\cup n_0$ random-mappings\;


Evaluate all mappings in {\it Anchors} by  {\it CloudSim}\label{alg:sample:a3}\;

\BlankLine
{\it Randoms} $\leftarrow N$ random mappings\label{alg:sample:gs}\;


\BlankLine
\ForEach{$r \in $ Randoms}{
$a_n \leftarrow$ mapping in {\it Anchors} that nearest to $r$\;
$a_f \leftarrow$ mapping in {\it Anchors} that furthest to $a_n$\;
\ForEach{$o \in\{\mathit{makespan}, \mathit{cost}\}$}{
\begin{spacing}{1.5}
$o_n, o_f \leftarrow o(a_n), o(a_f)$\;

$d_0, d_1 \leftarrow \mathit{dist}(a_n,r), \mathit{dist}(a_n,a_f)$\; 
$\theta \leftarrow \arccos(||\overrightarrow{a_nr}||/||\overrightarrow{a_na_f}||)$\;
$\widehat{o}_r \leftarrow \frac{d_0cos\theta}{d_1}\left(o_f-o_n\right) + o_n$\label{alg:sample:ge}\;

\end{spacing}
}
}

\BlankLine
$\{\mathcal{T}\} \leftarrow $ {\it Anchors} $\cup$ {\it Randoms}\;
\Return{$\mathcal{T}$}\;
\caption{Surrogate Based Evaluation}\label{alg:sample}
\end{algorithm}

%% file: tables/t_runtime.tex
\newcolumntype{M}{>{\centering\arraybackslash}m{2cm}}
\newcolumntype{G}{>{\centering\arraybackslash\columncolor{black!5}}m{0.65cm}}
\newcolumntype{L}{>{\centering\arraybackslash}m{1.2cm}}

\begin{table*}
\caption{Median Runtime* among 30 Repeats in RIOT and Others}
\footnotesize
\centering
\begin{threeparttable}
\begin{tabular}{c|GMMMMM|M}
    \toprule
    \scriptsize
    Model & MAKE SPAN* &RIOT ($t$) & MO-HEFT ($t_1$) &EMSC-NSGAII ($t_2)$ & EMSC-SPEA2 ($t_3$) & EMSC-MOEA
    /D ($t_4$) & Speedup $\frac{\min(t_i)}{t}$\\\toprule
    \footnotesize
 
M.25 & 31 & 19 & 6 & 33 & 34 & 35 & 0\\
M.50 & 49 & 9 & 17 & 62 & 65 & 65 & 1\\
M.100 & 113 & 6 & 55 & 154 & 157 & 162 & 9\\
M.1000 & 257 & 250 & 1.6H & 2.6H & 2.9H & 2.8H & 22\\
\hline
E.24 & 0.4H & 1 & 4 & 27 & 28 & 29 & 4\\
E.46 & 0.6H & 2 & 14 & 58 & 59 & 61 & 7\\
E.100 & 3.6H & 5 & 56 & 153 & 157 & 164 & 11\\
E.997 & 5.2H & 211 & 1.6H & 2.6H & 2.7H & 3.0H & 27\\
\hline
I.30 & 655 & 1 & 6 & 34 & 35 & 36 & 6\\
I.50 & 954 & 2 & 15 & 62 & 64 & 65 & 7\\
I.100 & 700 & 7 & 63 & 175 & 182 & 199 & 9\\
I.1000 & 0.5H & 189 & 1.5H & 2.6H & 2.7H & 4.8H & 29\\
\hline
C.30 & 124 & 1 & 7 & 34 & 35 & 36 & 7\\
C.50 & 198 & 3 & 16 & 63 & 63 & 65 & 5\\
C.100 & 241 & 7 & 61 & 154 & 154 & 162 & 8\\
C.1000 & 0.4H & 273 & 1.7H & 3.0H & 2.6H & 4.8H & 23\\
\hline
S.30 & 1006 & 1 & 6 & 34 & 35 & 35 & 6\\
S.60 & 1152 & 3 & 23 & 76 & 77 & 79 & 7\\
S.100 & 1133 & 9 & 62 & 168 & 169 & 180 & 6\\
S.1000 & 1.2H & 302 & 1.4H & 2.0H & 2.0H & 2.2H & 16\\

    \bottomrule
  \end{tabular}
 \begin{tablenotes}
\footnotesize
\item Runtime* is in seconds unless otherwise stated (H=hours).
\item Makespan* is the median makespan of all non-dominated scheduling found by any algorithm ran in the experiment.  
\item Model M/E/I/C/S = Montage, Epigenomics, Inspiral, CyberShake, Sipht (see \fig{structure})
\end{tablenotes}
\end{threeparttable}
 
\label{tab:runtime}
\end{table*}

%% file: tables/bigtable.tex
\sethlcolor{lightgray}
\newcommand{\sval}[2]{\bf{#1} {\footnotesize \it ({#2})}}
\newcommand{\ssval}[2]{{#1} {\footnotesize \it ({#2})}}
\newcolumntype{x}{>{\centering\arraybackslash\columncolor{black!5}}m{0.9cm}}
\begin{table*}
\caption{Median Measurements for all Experimented Workflows}
\centering
\footnotesize
\setlength{\tabcolsep}{0.41em}
\begin{threeparttable}
\begin{tabular}{l||xccccc|xccccc|xcccc}
\toprule
\multicolumn{1}{c||}{}& \multicolumn{6}{c|}{\bf Hypervolume}  &\multicolumn{6}{c|}{\bf IGD}&\multicolumn{5}{c}{\bf Spread}\\
 {\bf Model} &  RIOT &  MH &EN&ES&EM & RAND & RIOT  &  MH&EN&ES&EM & RAND & RIOT & MH&EN&ES&EM \\
\toprule
Montage 25 & \sval{79}{1} & 38 & 82 & 80 & 70 & 47 & \sval{4}{1} & 36 & 2 & 6 & 10 & 26 & \sval{83}{1} & 91 & 92 & 55 & 71\\
Montage 50 & \sval{82}{1} & 28 & 86 & 85 & 74 & 51 & \sval{4}{1} & 42 & 6 & 4 & 10 & 22 & \sval{76}{1} & ~{\it n.a.} & 105 & 62 & 77\\
Montage 100 & \sval{79}{1} & 29 & 85 & 83 & 77 & 49 & \sval{3}{1} & 48 & 2 & 8 & 22 & 30 & \sval{88}{1} & ~{\it n.a.} & 123 & 81 & 81\\
Montage 1000 & \ssval{75}{1} & 34 & 84 & 83 & 82 & 51 & \sval{3}{1} & 50 & 0 & 2 & 10 & 24 & \sval{96}{1} & ~{\it n.a.} & 104 & 55 & 93\\
\hline
Epigenomics 24 & \sval{73}{1} & 27 & 78 & 78 & 60 & 45 & \sval{8}{1} & 45 & 2 & 5 & 14 & 28 & \sval{95}{1} & 87 & 79 & 78 & 96\\
Epigenomics 46 & \sval{67}{1} & 0 & 68 & 68 & 62 & 28 & \sval{6}{1} & 67 & 5 & 12 & 17 & 35 & \sval{79}{1} & ~{\it n.a.} & 89 & 81 & 85\\
Epigenomics 100 & \sval{75}{1} & 6 & 70 & 69 & 64 & 35 & \sval{4}{1} & 62 & 5 & 3 & 22 & 31 & \ssval{93}{1} & 87 & 82 & 56 & 89\\
Epigenomics 997 & \sval{80}{1} & 0 & 68 & 67 & 65 & 31 & \ssval{7}{1} & 156 & 3 & 7 & 7 & 32 & \sval{60}{1} & 90 & 95 & 54 & 78\\
\hline
Inspiral 30 & \sval{72}{1} & 29 & 79 & 75 & 63 & 46 & \sval{6}{1} & 40 & 1 & 7 & 16 & 24 & \sval{95}{1} & 92 & 104 & 70 & 90\\
Inspiral 50 & \sval{72}{1} & 22 & 78 & 70 & 58 & 42 & \sval{4}{1} & 39 & 1 & 5 & 16 & 22 & \sval{93}{1} & 91 & 114 & 66 & 90\\
Inspiral 100 & \sval{69}{1} & 0 & 73 & 71 & 68 & 35 & \sval{4}{1} & 64 & 1 & 3 & 28 & 30 & \sval{80}{1} & ~{\it n.a.} & 109 & 79 & 71\\
Inspiral 1000 & \sval{80}{1} & 14 & 81 & 77 & 81 & 47 & \sval{4}{1} & 49 & 0 & 2 & 12 & 32 & \sval{84}{1} & 102 & 136 & 92 & 78\\
\hline
CyberShake 30 & \ssval{65}{1} & 37 & 82 & 81 & 67 & 33 & \ssval{14}{1} & 48 & 2 & 5 & 10 & 38 & \ssval{107}{1} & ~{\it n.a.} & 91 & 42 & 83\\
CyberShake 50 & \sval{67}{1} & 30 & 78 & 76 & 53 & 24 & \ssval{11}{1} & 43 & 2 & 5 & 16 & 41 & \ssval{108}{1} & 86 & 98 & 42 & 81\\
CyberShake 100 & \ssval{63}{1} & 19 & 74 & 72 & 61 & 13 & \sval{9}{1} & 55 & 3 & 5 & 16 & 55 & \ssval{110}{1} & 95 & 91 & 51 & 95\\
CyberShake 1000 & \sval{79}{1} & 39 & 81 & 81 & 80 & 49 & \sval{5}{1} & 38 & 1 & 2 & 13 & 30 & \sval{86}{1} & ~{\it n.a.} & 119 & 55 & 85\\
\hline
Sipht 30 & \sval{72}{1} & 28 & 74 & 74 & 23 & 10 & \sval{3}{1} & 38 & 1 & 3 & 38 & 89 & \ssval{111}{1} & 78 & 89 & 67 & 90\\
Sipht 60 & \ssval{71}{1} & 23 & 79 & 79 & 68 & 40 & \sval{7}{1} & 49 & 3 & 6 & 13 & 32 & \sval{96}{1} & 85 & 106 & 85 & 94\\
Sipht 100 & \ssval{68}{1} & 33 & 77 & 78 & 67 & 36 & \sval{5}{1} & 41 & 2 & 5 & 14 & 33 & \sval{89}{1} & ~{\it n.a.} & 103 & 85 & 88\\
Sipht 1000 & \ssval{68}{2} & 16 & 77 & 70 & 75 & 43 & \sval{5}{2} & 48 & 1 & 5 & 8 & 32 & \sval{94}{1} & ~{\it n.a.} & 127 & 79 & 71\\

\bottomrule
\end{tabular}
 \begin{tablenotes}\footnotesize
\item \textbf{\textit{Note: all values in the table are in $\mathbf{10^{-2}}$, e.g., the median hypervolume of Montage 25 gained from RIOT is $\mathbf{79*10^{-2}}$, i.e. 0.79}}
\item RAND = Random search (Sanity check)
\item MH = MOHEFT; EN = EMSC-NSGA-II;  ES = EMSC-SPEA2; EM = EMSC-MOEA/D
\item Hypervolume = higher are better; IGD = lower are better; Spread = lower are better
\item {\it\footnotesize (values)} next to RIOT are IQR (interquartile range, i.e. difference between 75th and 25th percentiles) of 30 repeats of RIOT
\item {\it n.a.} indicates no enough frontier points to calculate spread
\item {\bf Bold} values indicate that RIOT performed as well as or better than any of MH/EN/ES/EM (under Wilcoxon Test).
\end{tablenotes}
\end{threeparttable}
 
\label{tab:whole}
\end{table*}

%% file: main.bbl

%% file: main.bbl
\begin{thebibliography}{10}
\providecommand{\url}[1]{#1}
\csname url@samestyle\endcsname
\providecommand{\newblock}{\relax}
\providecommand{\bibinfo}[2]{#2}
\providecommand{\BIBentrySTDinterwordspacing}{\spaceskip=0pt\relax}
\providecommand{\BIBentryALTinterwordstretchfactor}{4}
\providecommand{\BIBentryALTinterwordspacing}{\spaceskip=\fontdimen2\font plus
\BIBentryALTinterwordstretchfactor\fontdimen3\font minus
  \fontdimen4\font\relax}
\providecommand{\BIBforeignlanguage}[2]{{%
\expandafter\ifx\csname l@#1\endcsname\relax
\typeout{** WARNING: IEEEtran.bst: No hyphenation pattern has been}%
\typeout{** loaded for the language `#1'. Using the pattern for}%
\typeout{** the default language instead.}%
\else
\language=\csname l@#1\endcsname
\fi
#2}}
\providecommand{\BIBdecl}{\relax}
\BIBdecl

\bibitem{vockler2011experiences}
V{\"o}ckler \emph{et~al.}, ``Experiences using cloud computing for a scientific
  workflow application,'' in \emph{Proceedings of the 2nd Intl. workshop on
  sci. cloud computing}.\hskip 1em plus 0.5em minus 0.4em\relax ACM, 2011, pp.
  15--24.

\bibitem{topcuoglu2002performance}
H.~Topcuoglu, S.~Hariri, and M.-y. Wu, ``Performance-effective and
  low-complexity task scheduling for heterogeneous computing,'' \emph{IEEE
  Trans. on parallel and distributed systems}, vol.~13, no.~3, pp. 260--274,
  2002.

\bibitem{rodriguez2017taxonomy}
M.~A. Rodriguez and R.~Buyya, ``A taxonomy and survey on scheduling algorithms
  for scientific workflows in iaas cloud computing environments,''
  \emph{Concurrency and Computation: Practice and Experience}, vol.~29, no.~8,
  2017.

\bibitem{chen2009ant}
W.-N. Chen and J.~Zhang, ``An ant colony optimization approach to a grid
  workflow scheduling problem with various qos requirements,'' \emph{IEEE
  Trans. on Systems, Man, and Cybernetics}, vol.~39, no.~1, pp. 29--43, 2009.

\bibitem{dorigo2006ant}
M.~Dorigo, M.~Birattari, and T.~Stutzle, ``Ant colony optimization,''
  \emph{IEEE computational intelligence magazine}, vol.~1, no.~4, pp. 28--39,
  2006.

\bibitem{rodriguez2014deadline}
M.~A. Rodriguez and R.~Buyya, ``Deadline based resource provisioningand
  scheduling algorithm for scientific workflows on clouds,'' \emph{IEEE Trans.
  on Cloud Computing}, vol.~2, no.~2, pp. 222--235, 2014.

\bibitem{shi2001particle}
Y.~Shi \emph{et~al.}, ``Particle swarm optimization: developments, applications
  and resources,'' in \emph{evolutionary computation. Proceedings of Congress
  on}, vol.~1.\hskip 1em plus 0.5em minus 0.4em\relax IEEE, 2001, pp. 81--86.

\bibitem{tsai2014hyper}
C.-W. Tsai \emph{et~al.}, ``A hyper-heuristic scheduling algorithm for cloud,''
  \emph{IEEE Trans. on Cloud Computing}, vol.~2, no.~2, pp. 236--250, 2014.

\bibitem{zhu2016evolutionary}
Z.~Zhu \emph{et~al.}, ``Evolutionary multi-objective workflow scheduling in
  cloud,'' \emph{IEEE Trans. on parallel and distributed Systems}, vol.~27,
  no.~5, pp. 1344--1357, 2016.

\bibitem{rehman2010initial}
M.~S. Rehman and M.~F. Sakr, ``Initial findings for provisioning variation in
  cloud computing,'' in \emph{Cloud Computing Technology and Science
  (CloudCom), 2010 IEEE Second Intl. Conference on}.\hskip 1em plus 0.5em minus
  0.4em\relax IEEE, 2010, pp. 473--479.

\bibitem{schad2010runtime}
J.~Schad, J.~Dittrich, and J.-A. Quian{\'e}-Ruiz, ``Runtime measurements in the
  cloud: observing, analyzing, and reducing variance,'' \emph{Proceedings of
  the VLDB Endowment}, vol.~3, no. 1-2, pp. 460--471, 2010.

\bibitem{iosup2011performance}
A.~Iosup, N.~Yigitbasi, and D.~Epema, ``On the performance variability of
  production cloud services,'' in \emph{Cluster, Cloud and Grid Computing
  (CCGrid), 2011 11th IEEE/ACM Intl. Sym. on}.\hskip 1em plus 0.5em minus
  0.4em\relax IEEE, 2011, pp. 104--113.

\bibitem{singh2017machine}
A.~Singh, A.~Rao, S.~Purawat, and I.~Altintas, ``A machine learning approach
  for modular workflow performance prediction,'' in \emph{Proceedings of the
  12th Workshop on Workflows in Support of Large-Scale Science}.\hskip 1em plus
  0.5em minus 0.4em\relax ACM, 2017, p.~7.

\bibitem{hsu2016inside}
C.-J. Hsu, R.~K. Panta, M.-R. Ra, and V.~W. Freeh, ``Inside-out: Reliable
  performance prediction for distributed storage systems in the cloud,'' in
  \emph{Reliable Distributed Systems, 2016 IEEE 35th Sym. on}.\hskip 1em plus
  0.5em minus 0.4em\relax IEEE, 2016, pp. 127--136.

\bibitem{durillo2014multi}
J.~J. Durillo and R.~Prodan, ``Multi-objective workflow scheduling in amazon
  ec2,'' \emph{Cluster computing}, vol.~17, no.~2, pp. 169--189, 2014.

\bibitem{nunez2012icancloud}
A.~N{\'u}{\~n}ez, J.~L. V{\'a}zquez-Poletti, A.~C. Caminero, G.~G.
  Casta{\~n}{\'e}, J.~Carretero, and I.~M. Llorente, ``icancloud: A flexible
  and scalable cloud infrastructure simulator,'' \emph{Journal of Grid
  Computing}, vol.~10, no.~1, pp. 185--209, 2012.

\bibitem{cai2017elasticsim}
Z.~Cai, Q.~Li, and X.~Li, ``Elasticsim: A toolkit for simulating workflows with
  cloud resource runtime auto-scaling and stochastic task execution times,''
  \emph{Journal of Grid Computing}, vol.~15, no.~2, pp. 257--272, 2017.

\bibitem{wang2017using}
Y.~Wang \emph{et~al.}, ``Using integer programming for workflow scheduling in
  the cloud,'' in \emph{Cloud Computing, 2017 IEEE 10th Intl. Conference
  on}.\hskip 1em plus 0.5em minus 0.4em\relax IEEE, 2017, pp. 138--146.

\bibitem{calheiros2011cloudsim}
R.~N. Calheiros \emph{et~al.}, ``Cloudsim: a toolkit for modeling and
  simulation of cloud computing environments and evaluation of resource
  provisioning algorithms,'' \emph{Software: Practice and experience}, vol.~41,
  no.~1, pp. 23--50, 2011.

\bibitem{hajela1990genetic}
P.~Hajela, ``Genetic search-an approach to the nonconvex optimization
  problem,'' \emph{AIAA journal}, vol.~28, no.~7, pp. 1205--1210, 1990.

\bibitem{aggarwal2001surprising}
C.~C. Aggarwal, A.~Hinneburg, and D.~A. Keim, ``On the surprising behavior of
  distance metrics in high dimensional spaces,'' in \emph{ICDT}, vol.~1.\hskip
  1em plus 0.5em minus 0.4em\relax Springer, 2001, pp. 420--434.

\bibitem{marler2004survey}
R.~T. Marler and J.~S. Arora, ``Survey of multi-objective optimization methods
  for engineering,'' \emph{Structural and multidisciplinary optimization},
  vol.~26, no.~6, pp. 369--395, 2004.

\bibitem{wang2016practical}
S.~Wang, S.~Ali, T.~Yue, Y.~Li, and M.~Liaaen, ``A practical guide to select
  quality indicators for assessing pareto-based search algorithms in
  search-based software engineering,'' in \emph{Software Engineering (ICSE),
  2016 IEEE/ACM 38th Intl. Conference on}.\hskip 1em plus 0.5em minus
  0.4em\relax IEEE, 2016, pp. 631--642.

\bibitem{Wolpert97}
D.~H. Wolpert and W.~G. Macready, ``No free lunch theorems for optimization,''
  \emph{IEEE Trans. on Evolutionary Computation}, vol.~1, no.~1, pp. 67--82,
  Apr 1997.

\end{thebibliography}
